\documentclass[12pt]{article}
\usepackage{cite}
\usepackage{amsmath,amsthm, amscd, amssymb, amsfonts}
\usepackage{appendix}
\usepackage{graphicx}
\begin{document}
\title{A Note on Bell's Theorem Logical Consistency}
\author{Justo Pastor Lambare and Rodney Franco}
\date{}
\maketitle
\begin{abstract}
Counterfactual definiteness is supposed to underlie the Bell theorem.
An old controversy exists among those who reject the theorem implications by rejecting counterfactual definiteness and those who claim that, since it is a direct consequence of locality, it cannot be independently rejected.
We propose a different approach for solving this contentious issue by realizing that counterfactual definiteness is an unnecessary and inconsistent assumption.
Counterfactual definiteness is not equivalent to realism or determinism neither it follows from locality.
It merely reduces to an incongruent application of counterfactual reasoning.
Being incompatible with falsifiability, it constitutes an unjustified assumption that goes against the scientific method rigor.
Correct formulations of the Bell theorem's bases show it is absent either as a fundamental hypothesis or as a consequence of something else.
Most importantly, we present a coherent Bell inequality derivation carefully devised to show explicitly and convincingly the absence of incompatible experiments or counterfactual reasoning.
Thus, even admitting that counterfactual definiteness could be a consistent assumption, the necessary conclusion is that it is irrelevant for the inequality formulation and can be safely ignored when discussing Bell's inequality philosophical and physical implications.
\end{abstract}


\tableofcontents  
\section{Introduction}\label{sec:INTRO}
Some persistent confusions concerning the \emph{Clauser, Horne, Shimony, and Holt} (CHSH) \cite{pCHSH69} form the the \emph{Bell inequality} (BI) jeopardize Bell theorem's logical soundness and scientific relevance.
We critically analyze two different positions involving similar inconsistencies.

An orthodox view has produced unnecessary confusion and unjustified criticisms, often leading to incorrect conclusions.
It arises as a consequence of an extraneous \emph{ad hoc} assumption: \emph{counterfactual definiteness} (CFD). Moreover, another unorthodox stance, somehow related to the former, based on the existence of joint probabilities.

Since CFD's validity is the widely accepted view, we shall mainly concentrate on it.
CFD can be defined as \emph{``the ability to speak ``meaningfully'' of the definiteness of the results of measurements that have not been performed''} \cite{wiki:cfd}.

We shall see that ``meaningfully'' in the above definition signifies giving meaning to what, otherwise, is devoid of physical and logical sense.
Whereas it is not explicit in its definition, CFD concedes that it is possible to obtain valid conclusions comparing unrelated results.
This tacit assumption constitutes its \emph{raison d'\^{e}tre}, CFD being a euphemism to smuggle it as a realism\footnote{Whatever the meaning of realism one would like to adopt, see for instance Ref.\cite{pNor07}.}  assumption.

Notice that CFD is more than just \emph{counterfactual reasoning}(CR).
Since their difference may appear fuzzy, and they are not normally distinguished, we explicitly state their difference:
although both speculate about counterfactual predictions, CFD further assumes those predictions are falsifiable, notwithstanding their irreproducibility.

Admittedly, determinism allows the correct prediction of would be impossible experiments.
However, to assume real experiments can falsify theoretical results which reproducibilities are unwarranted by experimental protocols is against the tenets of the scientific falsification method.
The mere declared definiteness of those imaginary results does not alleviate the irreproducibility problem; therefore, their testability should be rigorously justified.

CFD became widespread since H. P. Stapp first introduced CR \cite{pSta71} to prove the Bell theorem without hidden variables.
Stapp, however, did not derive a Bell-type inequality; he obtained a mathematical contradiction instead of a directly falsifiable inequality.
Despite often claims to the contrary, neither CHSH nor John Bell ever mentioned ``not performed'' or ``incompatible'' experiments in their proofs.
So it is justified to assume they did not apply CFD and that those claims are based on incorrect interpretations of their calculations.
Henceforth we shall use the acronym CFD-BI for the BI interpreted according to CFD.

The CFD-BI generated a long-standing interpretational debate \cite{pMau10,pLau18a,pSky82,pFor86,pDic93,pZuk14,pCza20,pLam20b} over the misgivings produced by the use of subjunctive conditionals to prove the BI.
Some consider CFD an independent hypothesis \cite{pPer78,pBla10}.
Others view CFD as equivalent or implied by realism \cite{pGil14,pBou17}.
A third stance equates CFD to determinism, considering it a ``harmless'' consequence of locality.
The last point is important.
Often criticisms involving CFD are rejected on the basis that it is a logical consequence of locality  \cite{pMau10,pLau18a}.
The main objective of our work is to present definitive proof of CFD irrelevance regarding the validity and implications of the BI by introducing a derivation that clearly excludes any explicit or implicit reference to CFD.

John Bell's keen insight allowed him to take the \emph{Einstein, Podolsky, and Rosen} (EPR) \cite{pEPR35} argument a step further, replacing a thought experiment with a testable theoretical prediction.
Bell not only managed to come up with a numerical prediction but -- and this is the generally overlooked essential point -- conceived it through a realizable and meaningful physical experiment\footnote{Still, experimentalists had to refine it to make it more appropriate for the experimental tests \cite{pCHSH69}.}.

The paper is organized in the following way.
Section \ref{sec:THUBDM} presents the minimal reasonable assumptions necessary to establish Bell's deterministic model of hidden variables, pointing out the absence of any form of CR.
This section is relevant; it shows that we can obtain determinism without CR, unlike the EPR argument, which is usually considered based on CR.
Section \ref{sec:ICDF} explains the historical origin of CR as applied to the BI and why CFD is an inconsistent assumption.
Readers interested only in reviewing a BI derivation that avoids the CFD assumption may skip this section altogether without loss of continuity.
The central part of the paper is section \ref{sec:DERIV}.
There we introduce an unprecedented BI derivation painstakingly conceived to show CFD's absence either as a fundamental assumption or as a derived consequence.
The derivation is unprecedented in the sense that, by explicitly excluding CFD, it leaves no room for infiltration of doubts about the absence of counterfactual reasoning.
Of course, we firmly believe that John Bell and CHSH never meant otherwise.
Finally, section \ref{sec:JP} critically reviews some bizarre claims related to joint probabilities.
\section{The Hypotheses Underlying BI}\label{sec:THUBDM}
Often Bell's deterministic model of \emph{Local Hidden Variables} (LHV) is presented directly as comprising the functions $A(a,\lambda)$, $B(b,\lambda)$, and the probability distribution $p(\lambda)$.
On the other hand, from a conceptual viewpoint and to avoid misinterpretations, it is convenient to derive Bell's model from two fundamental assumptions: \emph{Local Causality}(LC) and \emph{Measurement Independence}(MI).
LC and MI are concrete and physically motivated concepts amenable to a mathematical formulation.
No other assumptions are necessary for a coherent BI derivation, as we shall prove in section \ref{sec:DERIV}.

Local causality requires starting with a probabilistic approach.
Let $P(A,B|a,b)$ be Alice and Bob's conditional joint probability of obtaining $A$ and $B$ as results of their measurements when their settings are $a$ and $b$.

If $\lambda$ represents the common causes lying in the intersection of their respective past light cones, then, according to \emph{Reichenbach's Principle of Common Causes}(RPCC) \cite{RPCC}\footnote{Bell did not mention Reichenbach, but his local hidden variables play the same role as Reichenbach's common causes.}, LC requires that
\begin{equation}\label{eq:lc}
P(A,B|_{a,b,\lambda})=P(A|_{a,\lambda}) P(B|_{b,\lambda})
\end{equation}
Where the first and second factors in the RHS of (\ref{eq:lc}) are Alice and Bob's respective probabilities.
Assuming \emph{perfect correlations}(PC) and putting $A=B$ and $a=b$ in (\ref{eq:lc}), either $P(A|a,\lambda)=0$ or $P(B|b,\lambda)=0$; if $P(A|a,\lambda)=0$ then $P(-A|a)=1$ and analogously for $P(B|b,\lambda)$.

Since $A,\,B\,,a,$ and $b$ represent any values in their respective domains, we have Alice's and Bob's probabilities reduces to zeros and ones.
Thus, (\ref{eq:lc}) and PC imply determinism(D) \cite{pFra82,pHal20}
\begin{equation}\label{leq:D}
LC\wedge PC\rightarrow D
\end{equation}
The transition from the stochastic function $P(A|_{a,\lambda})\in\{0,1\}$ to the deterministic function $A(a,\lambda)\in\{-1,+1\}$ is given by:
\begin{equation}\label{eq:dtm}
A(a,\lambda)=\left\{
\begin{array}{cc}
+1  & ,when \, P( 1|_{a,\lambda})=1\\
-1  & ,when \, P( 1|_{a,\lambda})=0
\end{array}
\right.
\end{equation}
Similarly, $B(b,\lambda)$ is defined.
Determinism arises without CR or CFD.
The formulation is in terms of indicative conditionals; subjunctive conditionals do not appear.
Please also notice that we do not claim to have derived CFD as a consequence of determinism.
The use of counterfactuals is philosophically problematic, physically unconvincing \cite{bVai09}, and, as we argue in section \ref{sec:ICDF}, experimentally inconsistent.

In section \ref{sec:DERIV}, we show it is possible to derive the BI by predicting only the results of experiments that are supposed to be actually performed; we do not need to assume that results exist before measurement.
The existence of \emph{EPR elements of physical reality} is a stronger and unnecessary assumption.
Einstein famously disliked that metaphysical idea.
He preferred to formulate the quantum mechanics incompleteness argument through his separation principle instead \cite{pHow85}.\footnote{When we assume the pre-existence of elements of physical reality as a manifestation of realism, we can prove CFD inconsistency by more direct methods, see Ref. \cite{pLam19}.}
The last point is important.
We often hear that the BI is violated because elements of physical reality do not exist.
Since the correct BI formulation only predicts the results of experiments that are supposed to be actually performed, it is neutral regarding whether they exist before measurement or are created by measurement.\footnote{In Ref. \cite{bBel71}, Bell explicitly considered the possibility that measurement results be the consequence of interactions with the measuring devices by introducing apparatuses' hidden variables.}

MI means that $p(\lambda)$ is independent of the experimental choices.
It is physically justified by requiring the setting variables to be uncorrelated with the common causes $\lambda$.
It is also known as the no conspiracy assumption.
Let $p(ab)$ be the joint probability of settings $a$ and $b$ occurrence, then ``no conspiracy'' requires that
\begin{equation}\label{eq:bayes1}
p(ab|_\lambda)=p(ab)
\end{equation}
The LHS is the probability conditional on the common causes that determine the spin measurement results.
According to Bayes's theorem of probability theory
\begin{equation}\label{eq:bayes2}
p(ab\lambda)=p(ab|_\lambda)p(\lambda)=p(\lambda|_{ab}) p(ab)
\end{equation}
From (\ref{eq:bayes1}) and (\ref{eq:bayes2}) we obtain MI
\begin{equation}\label{eq:mi}
p(\lambda)=p(\lambda|_{ab})
\end{equation}
\section{Implications of the CFD Hypothesis}\label{sec:ICDF}
In this section, we analyze CFD's consequences for the BI scientific and objective meaning.
Anticipating that many readers would not accept CFD inconsistency, some considering it a hallmark of ``classicality'' while others a consequence of locality, those readers may want to jump to section \ref{sec:DERIV}, where we present a Bell inequality derivation that explicitly and clearly excludes any form of counterfactual reasoning.
That derivation puts it abundantly clear that even if we assume CFD either as a derived or fundamental consistent principle, it is, nonetheless, irrelevant concerning BI validity.

Here we explain why CFD is different from determinism or realism and does not follow from locality.
In section \ref{ssec:IFIF}, we go over CFD implications for the inequality's falsifiability.
In section \ref{ssec:TCBITE}, we analyze the CFD-BI as a thought experiment comparing it with the \emph{Greenberger-Horne-Zeilenger} (GHZ) Theorem \cite{pGHZ90} also known as Bell's theorem without inequalities.
\subsection{Implications for the BI's Falsifiability}\label{ssec:IFIF}
CFD's consequences for the inequality falsifiability pass unnoticed owed to the disregard for its experimental implementation, overlooking the difference between predictability and testability.
Of course, thought experiments are useful for gaining insight into many situations.
However, when they are impossible to implement, comparing their results with those of actual experiments should be careful.
Not every imaginable result that our theory predicts can be put to the test in real experiments.

As we shall review in section \ref{sec:DERIV}, BI's derivation requires considering the following expression containing eight measurement results with the same value of hidden variable $\lambda$ (see eq. (\ref{eq:cs}) in sect. \ref{sec:DERIV})
{\footnotesize
\begin{equation}\label{eq:csa}
 A(a_1,\lambda)B(b_1,\lambda)-A(a_1,\lambda)B(b_2,\lambda)+A(a_2,\lambda)B(b_1,\lambda)+A(a_2,\lambda)B(b_2,\lambda)
\end{equation}
}
When we conceive the realization of (\ref{eq:csa}) through counterfactual experiments,  H. P. Stapp \cite{pSta71} eloquently put it: \emph{``Of these eight numbers only two can be compared directly to experiment. The other six correspond to the three alternative experiments that could have been performed but were not''}.

The former interpretation of (\ref{eq:csa}) bears no relation with the data collected in real experiments.
In CHSH experiments, entangled pairs are created then measured in Alice's and Bob's detectors.
There is no way to relate a given measurement with the result a previous experiment would have produced if it was performed with a different setting.
Hence, the results of real experiments cannot be meaningfully compared with the theoretical predictions when we derive the inequality interpreting (\ref{eq:csa}) as containing counterfactual outcomes.
In other words, no known theoretical principle or experimental evidence exists that justifies interpreting actual data according to (\ref{eq:csa}).
This happens because experimentalists have no control over the hidden variables.
Indeed, we do not even know if they exist!
On the contrary, the experiment is supposed to falsify their existence.
Although we can assume their existence, we cannot further assume the same values of hidden variables will appear in other experiments performed with different setting combinations forming groups of perfectly convenient sets as in (\ref{eq:csa}) so that the inequality be bounded by 2.
At least, the assumption that they are counterfactually definite does not validate their materialization out of thin air.
In section \ref{sec:DERIV}, assuming just LC and MI, we show that (\ref{eq:csa}) indeed arises naturally with actual experimental data.

A possible solution for validating (\ref{eq:csa}), when containing counterfactual results, is to consider the counterfactual terms as a calculational artifice, hence preventing the need for their experimental replications.
Although it is possible to add hypothetical terms in a theoretical derivation without altering the final result, in these cases, what is ``artificially'' added has to amount to zero.
That does not happen when we add in (\ref{eq:csa}) three counterfactual terms to the actual one.
Three values $\pm1$ cannot add to zero. Hence, the addition of the counterfactual terms alters the result that is supposed to be found in the experiment  \cite{pLam20b}.

In an intention to make sense out of CFD, some suggest interpreting it statistically \cite{wiki:cfd}.
We can hypothesize that CFD allows the reproduction of counterfactual results with actually performed experiments through some unknown mechanism that would yield the desired statistical result.
However, such an unjustified ansatz goes against the usual rigor characterizing a factual science.
As Asher Peres wittily concluded after deriving a CFD-BI, ``unperformed experiments have no results'' \cite{pPer78}, which is tantamount to saying that ghosts do not exist.

With a few notable exceptions,\footnote{Hess and Philipp \cite{pHes04} refute the Bell theorem for different reasons but explain the CFD problem correctly, at least for the special case when the different settings occur with equal probability.} those noticing the inconsistency assume that it is an intrinsic problem of the Bell inequality, blaming John Bell for the blunder \cite{pCza20,bAde01,pJCh19,pJCh18,pDBa84a} that, as far as we know, started in the late 1970s\cite{pPer78,pEbe77}.

In our opinion, two articles rejecting the Bell theorem by correctly explaining the problem are Refs. \cite{bAde01} and \cite{pDBa84a}.
Although many refutations of the Bell theorem do not deserve much attention, the previously cited articles elicit valid criticisms that were not properly responded to.
This article can be considered a belated proper response.
\subsection{The CFD-BI as a Thought Experiment}\label{ssec:TCBITE}
Here we shall see the CFD-BI fails even as a thought experiment.
Besides being experimentally unstable, the CFD-BI cannot be meaningfully compared with the quantum mechanical prediction.

For that purpose, it will be useful to compare the CFD-BI with the GHZ theorem.
The comparison is useful because the last theorem presents a counterfactual result that is not directly testable but can be meaningfully compared with quantum mechanical predictions.

We succinctly review GHZ's argument. They consider four spin-1/2 particles in the entangled state
\begin{equation}\label{eq:GHZstate}
|\psi\rangle=\frac{1}{\sqrt{2}}[|+\rangle_1|+\rangle_2|-\rangle_3|-\rangle_4 - |-\rangle_1|-\rangle_2|+\rangle_3|+\rangle_4]
\end{equation}
Those four particles are sent to four different stations where spin measurements are performed under angles $a,b,c$, and $d$ measured with respect to a common direction lying in parallel planes which are orthogonal to the direction of motion.

Let $A, B, C$, and $D$ be the results $\pm1$ of these measurements. It can be shown that quantum mechanics predicts \cite{pGHZ90}
\begin{equation}
\langle ABCD\rangle=-\cos(a+b-c-d)
\end{equation}
Then we have perfect correlations for certain angles
\begin{eqnarray}
ABCD &=&-1,\qquad for\quad a+b-c-d=0 \label{eq:GHZ-1}\\
ABCD &=&+1,\qquad for\quad a+b-c-d=\pi \label{eq:GHZ+1}
\end{eqnarray}
although quantum mechanics dos not predict with certainty the results of the individual measurements $A, B, C, D$.

Through an ingenious reasoning GHZ proceed to prove that (\ref{eq:GHZ-1}) and (\ref{eq:GHZ+1}) are incompatible with the assumption of LHV, i.e., local deterministic functions $A(a,\lambda), B(b,\lambda)$,$C(c,\lambda)$, and $D(d,\lambda)$ cannot consistently reproduce the quantum mechanical predicted results (\ref{eq:GHZ-1}) and (\ref{eq:GHZ+1}).
It is interesting to note that GHZ's reasoning does not require the no conspiracy assumption.

To prove the above mentioned incompatibility, GHZ consider four different setting combinations -- constraint by $a+b-c-d=0$ -- on the same experimental run.
Since only one of them can actually be performed, the other three are counterfactual, i.e., three of the following equations represent results of experiments that could have been performed but were not
\begin{equation}\label{eq:GHZcf}
\begin{array}{ccc}
A(0,\lambda)B(0,\lambda)C(0,\lambda)C(0,\lambda)&=&-1\\
A(\phi,\lambda)B(0,\lambda)C(\phi,\lambda)C(0,\lambda)&=&-1\\
A(\phi,\lambda)B(0,\lambda)C(0,\lambda)C(\phi,\lambda)&=&-1\\
A(2\phi,\lambda)B(0,\lambda)C(\phi,\lambda)C(\phi,\lambda)&=&-1
\end{array}
\end{equation}
We do not need more details of the GHZ argument.
We only want to show why GHZ constitutes a valid thought experiment based on correct CR that proves the incompatibility of LHV with quantum mechanics.

Notice the analogy between (\ref{eq:csa}) and (\ref{eq:GHZcf}). In (\ref{eq:GHZcf}), only one equation correspond to the results of an actual experiment, the other three correspond to the results of three experiments that could have been performed but were not, just as the terms in (\ref{eq:csa}).
In both cases, CR is necessary to justify the same hidden variable $\lambda$ in the four different terms (\ref{eq:csa}) or equations (\ref{eq:GHZcf}).

Why counterfactual reasoning works for GHZ and fails for the BI?
In the GHZ case, we have that both theories -- quantum mechanics and LVH -- make predictions for the three counterfactual experiments in (\ref{eq:GHZcf}).
Therefore, it is meaningful to compare their predictions.
That comparison proves the incompatibility of both theories, notwithstanding the experimental irreproducibility of those counterfactual predictions.\footnote{To complete their derivation GHZ need to use (\ref{eq:GHZ+1}) obtaining a fifth counterfactual relation. However, this is not relevant for our discussion.}

In the case of the CFD-BI, quantum mechanics does not predict the counterfactual terms corresponding to a single experimental run\footnote{Precisely GHZ's ingenious trick was to figure out a configuration allowing quantum mechanics' deterministic predictions.}
{\scriptsize
\begin{equation}\label{eq:csqm}
A(a_1,b_1)B(a_1,b_1)-A(a_1,b_2)B(a_1,b_2)+A(a_2,b_1)B(a_2,b_1)+A(a_2,b_2)B(a_2,b_2)
\end{equation}
}
Therefore, we cannot compare (\ref{eq:csa}) with (\ref{eq:csqm}). Quantum mechanics only predicts the expectation values
\begin{equation}
\langle A(a_i,b_k)B(a_i,b_k)\rangle = \langle A_iB_k\rangle_{qm}
\end{equation}
Although we can take mean values in (\ref{eq:csa}) to obtain $\langle A_iB_k\rangle_{lhv}$, then have the quantum mechanical and the LHV predictions
\begin{eqnarray}
S_{qm} &=& \langle A_1B_1\rangle_{qm} -\langle A_1B_2\rangle_{qm} +\langle A_2B_1\rangle_{qm} +\langle A_2B_2\rangle_{qm}\\
S_{lhv} &=& \langle A_1B_1\rangle_{lhv} -\langle A_1B_2\rangle_{lhv} +\langle A_2B_1\rangle_{lhv} +\langle A_2B_2\rangle_{lhv}
\end{eqnarray}
It is not legitimate to compare $S_{qm}$ with $S_{lhv}$ since we obtained them under different conditions: three of the four terms in (\ref{eq:csqm}) are not warranted to represent the result we would have obtained if performed instead of the actually measured one.
Although both theories give different results, they do it under different conditions.

Thus, we cannot infer any valid conclusion by comparing their predictions.
To assume the difference of those conditions disappears because we take mean values is unjustified and solving the problem by postulating their equivalence is not cogent.
CFD is the dubious assumption that we can make this bogus comparison and still obtain a meaningful conclusion.
That goes against the usual rigor characterizing the hard sciences.
Hence, the counterfactual prediction under LHV $S_{lhv}$  -- apart from being unfalsifiable -- is also meaningless as a thought experiment.
\section{Making Sense of Bell's Derivation}\label{sec:DERIV}
Here we scrutinize the derivation making explicit CDF's absence, either as a fundamental assumption or derived consequence of something else.

We eliminate CFD  and remark that we predict only outcomes of experiments that are supposed to be actually performed without imposing irreproducible conditions, \emph{``That the experiment is going to be done at all, is of course an independent point; what we are meant to explain causally is that the outcome is thus and so if the experiment is done''}\cite{pFra82}.

In other words, we shall prove that an expression like (\ref{eq:csa}) is a ``natural'' consequence of the fundamental hypotheses LC and MI.
In the debate of ``intelligent design'' vs. ``Darwinian evolution'', it became obvious that science is the search for natural explanations.
It is somehow ironic that, in the physical sciences, we have to insist on a natural explanation over superstitious beliefs in materializations of convenient results. Such lack of rigor is improper of a factual science.

Thus, we shall present the steps followed in the derivation, highlighting how actual experimental results relate to LHV's predictions in a concrete and unambiguous form.
This shows that, even if we believe that some unspecified mechanism allows incompatible experiments to materialize out of thin air during the realization of real experiments, such an act of faith is not necessary to falsify the Bell inequality.
\subsection{The Derivation}\label{ssec:TDERIV}
Let $A_l(a_i,b_k)$ and $B_l(a_i,b_k)$ be the actual ``clicks'' detected by Alice's and Bob's joint measurement during the experiment's $l^{\underline{th}}$ run.
The experimental correlation is
{\small
\begin{equation}\label{eq:ect}
    E(a_i,b_k) = \frac{1}{N}\sum_{l=1}^N A_l(a_i,b_k)B_l(a_i,b_k)\,;\quad i,k\in\{1,2\}
\end{equation}
}
After the whole experiment has been run, we end up with four different experimentally measured correlations $E(a_1,b_1),\,E(a_1,b_2),\,E(a_2,b_1),\,E(a_2,b_2)$.
By adequately adding them
\begin{eqnarray}
S^*    &=& E(a_1,b_1)-E(a_1,b_2)+E(a_2,b_1)+E(a_2,b_2)\label{eq:eS}
\end{eqnarray}
The theoretical analysis of the experimental result $S^*$ presents two sides.\footnote{There is also the statistical analysis concerning the experimental errors and confidence intervals, which we shall not consider; our analysis being purely theoretical.}
One arises when we ask about the quantum mechanics' prediction for $S^*$.
The other appears when we ask what the LHV prediction is.
We are concerned exclusively with the last case.

The conflation of those different sides results in much confusion about the Bell theorem interpretation. 
When analyzing the LHV prediction, there are no questions about Hilbert spaces' non-commuting operators or observables' eigenvalues.
\emph{The LHV prediction exclusively concerns whether the functions $A(a,\lambda)$, $B(b,\lambda)$, and hidden variables with probability distribution $p(\lambda)$ can explain what has been experimentally found in four different series of actual experiments}.
There are no issues related to unperformed or incompatible experiments, joint probabilities, elements of physical reality, or metaphysical assumptions of any sort.
The problem under investigation and the proposed Bell's deterministic model are so simple and straightforward that people seem suspicious that such a stunning simplicity could have profound foundational consequences \cite{pNie09}.

Assuming the local deterministic functions $A(a,\lambda)$ and $B(b,\lambda)$ can describe the experimental ``clicks''
\begin{equation}\label{eq:fa}
\begin{array}{c}
A_l(a_i,b_k)=A(a_i,\lambda_l^*)\\
B_l(a_i,b_k)=A(b_k,\lambda_l^*)
\end{array}
\end{equation}
for some value $\lambda=\lambda_l^*$ of hidden variable. Taking (\ref{eq:fa}) in (\ref{eq:ect})
{\small
\begin{equation}\label{eq:ct1}
    E(a_i,b_k) = \frac{1}{N}\sum_{l=1}^N A_l(a_i,\lambda_l^*)B_l(b_k,\lambda_l^*)
\end{equation}
}
Associating terms with equal hidden variables' values in the RHS of (\ref{eq:ct1}) and taking the limit for $N\rightarrow\infty$
{\small
\begin{equation}\label{eq:ct2}
\sum_{j\in I} p(\lambda_j)\,A(a_i,\lambda_j)B(b_k,\lambda_j)=\lim_{N\rightarrow\infty}\frac{1}{N}\sum_{l=1}^N A(a_i,\lambda^*_l)B(b_k,\lambda^*_l)
\end{equation}
}
Where  $I$ is an index set that characterizes the hidden variables' domain; $\lambda^*_l\in\{\lambda_j: j\in I\}$.
In the LHS, $p(\lambda_j)$ is the relative frequency of $\lambda_j$, i.e., it could be obtained by counting the number of times $N_j$ a particular hidden variable's value $\lambda_j$ appeared in the RHS, and dividing it by $N$.\footnote{It is the frequentist interpretation of the distribution function $p(\lambda)$.} Measurement independence allows us to write $p(\lambda_j)$ instead of $p(\lambda_j|a_i, b_k)=p_{\,ik}(\lambda_j)$
\begin{equation}\label{eq:expro}
p(\lambda_j)=\lim_{N\rightarrow\infty}\frac{N_j}{N}
\end{equation}
Consequently, assuming LHV, the experimental correlations can be expressed as in the LHS of (\ref{eq:ct2})\footnote{In our analysis, we do not consider the problem of finite statistics \cite{pLar14}, so we shall ignore the limits in (\ref{eq:ct2}) and (\ref{eq:expro}).} and the RHS of (\ref{eq:eS}) becomes
%
\begin{eqnarray}
S &=&\sum_{j\in I} p(\lambda_j)\,A(a_1,\lambda_j)B(b_1,\lambda_j)-\sum_{j\in I} p(\lambda_j)\,A(a_1,\lambda_j)B(b_2,\lambda_j)\nonumber\\
  &+&\sum_{j\in I} p(\lambda_j)\,A(a_2,\lambda_j)B(b_1,\lambda_j)+\sum_{j\in I} p(\lambda_j)\,A(a_2,\lambda_j)B(b_2,\lambda_j)\label{eq:s1}
\end{eqnarray}
%
Since, per MI, the four sums in (\ref{eq:s1}) range over the same hidden variables' domain
\begin{eqnarray}
S    &=& \sum_{j\in I} p(\lambda_j)\,C(\lambda_j)\label{eq:s2}\\
|S|  &\leq& \sum_{j\in I} p(\lambda_j)\,|C(\lambda_j)|\label{eq:s3}\\
     &\leq& \sum_{j\in I} p(\lambda_j)\, 2\label{eq:s4}\\
     &\leq& 2\sum_{j\in I} p(\lambda_j)\label{eq:s5}\\
     &\leq& 2\label{eq:s6}
\end{eqnarray}
%
Where the total probability $\sum p(\lambda_j)=1$ and $C(\lambda_j)$ is given by
\begin{eqnarray}
  C(\lambda_j) &=& A(a_1,\lambda_j)B(b_1,\lambda_j)-A(a_1,\lambda_j)B(b_2,\lambda_j)+A(a_2,\lambda_j)B(b_1,\lambda_j)\nonumber\\
             &=&+A(a_2,\lambda_j)B(b_2,\lambda_j)\label{eq:cs} 
\end{eqnarray}
\emph{None of the terms present in (\ref{eq:cs}) are assumed to have originated from incompatible experiments, neither materialized out of counterfactual reasoning nor pre-existed before actually measured.}
$C(\lambda_j)$ emerges when we assume the real experimental data have the form given in  (\ref{eq:s1}), according to the hypotheses of Bell's LHV model.
The origin of those four terms can be traced back to the actual experimental ``clicks'' through (\ref{eq:ect}), (\ref{eq:fa}), (\ref{eq:ct1}), and (\ref{eq:ct2}).

Thus, if LHV is correct, (\ref{eq:cs}) contains only results of experiments that are supposed to have been actually performed.\footnote{Similar rational explanations for (\ref{eq:cs}) can be found in Refs. \cite{pLam20b,pLam17a}.}
\emph{That is the rational down to earth meaning of (\ref{eq:cs}) and the reason why real experiments falsify LHV's theoretical prediction.}
In an intention to bring back common sense to the field, Bas C. Van Fraassen rightfully observed: \emph{``A reader as yet unfamiliar with the literature will be astounded to see the incredible metaphysical extravaganzas to which this subject has led''}\cite{pFra82}.

Of course, since we are falsifying the LHV model, it may be the case (\ref{eq:cs}) might not actually happen.
Considering the above derivation predicts for $|S|$ an upper bound value of 2, and the experiments yield an actual value $|S^*|>2$, we must conclude, a posteriori, that at least one hypothesis assumed by the LHV model must be false and that (\ref{eq:cs}) does not occur after all.

Next, we analyze and close a possible loophole in the derivation we have made above.
\subsection{A Possible Loophole}\label{ssec:APL}
The climax of the above derivation is the attainment of (\ref{eq:cs}).
Without it, the upper bound 2 for $|S|$ is not warranted.
So, if we want to avoid BI implications -- i.e., LHV's untenability and its nonlocal implications --, all we have to do is find reasons to doubt the realization of (\ref{eq:cs}) notwithstanding LHV's correctness

There seems to be a loophole in our reasoning that a LHV advocate might exploit \cite{bAde01,pDBa84a}.
To accomplish (\ref{eq:cs}) with actual experimental results, we would need an exact repetition of hidden variables values $\lambda_j$ in four different series of experiments. Notice that each term in (\ref{eq:cs}) belongs to a differen series of experiments performed with different settings.

Since the hidden variables are unknown parameters that could mean anything, their exact values' repetition in different finite series of experiments seems highly improbable.
It would be justified only in a purely theoretical derivation.
For instance, they could belong to a continuous spectrum that would make their exact repetition by actual different experiments virtually impossible.

The fact that we need such exact repetitions of hidden variables in four different series of unrelated experiments seems to be a new independent and dubious hypothesis necessary to falsify LHV.
Willy De Baere\cite{pDBa84a} dubbed this assumption the \emph{Reproducibility Hypothesis}(RH) and, ironically, it seems to be nearly as implausible and unjustified as CFD's testability, which we are trying to avoid.
Next, we explain why the RH is not a necessary independent hypothesis and, if the LHV hypotheses hold, (\ref{eq:cs}) can represent actual experimental data.

Eugene Wigner \cite{pWig70} was the first to point out that only a small finite number of ``effective'' hidden variables exists in a Bell-type inequality test.
In the experiment that Wigner conceived, eight different hidden variables exist.

In a CHSH experiment, there are 16 different such variables \cite{pFin82}.
This does not mean the actual number of hidden variables depends on the experiment. It means that they behave as if there were only a small ``effective'' number according to the particular experimental configuration.

In the case of a CHSH experiment, if we set $A(a_i,\lambda)=A_i(\lambda)$, $B(b_k,\lambda)=B_k(\lambda)$ and define $\lambda_1\equiv\lambda_2$ if and only if
{\small
\begin{equation}\label{eq:eqclass}
(\,A_1(\lambda_1),A_2(\lambda_1),B_1(\lambda_1),B_2(\lambda_1)\,)=(\,A_1(\lambda_2),A_2(\lambda_2),B_1(\lambda_2),B_2(\lambda_2)\,)
\end{equation}
}
Then, at most, sixteen different equivalent classes of ``effective'' hidden variables appear in a given CHSH experiment.

Let $\overline{\lambda}_j$ represent an equivalent class as defined by (\ref{eq:eqclass}) and express all relevant magnitudes as functions of these equivalent classes.
When we perform more than 16 runs of experiments to obtain $A(a_i,\overline{\lambda}_j)B(b_k,\overline{\lambda}_j)$\footnote{With the obvious meaning $A(a_i,\overline{\lambda}_j)=A(a_i,\lambda_r)\,,\,\lambda_r\in\overline{\lambda}_j$.}, the hidden variables' classes will start to repeat, although the $\lambda$'s actual values need not be repeated.

When the Bell test experiment consists of a statistically significant number of runs, it will produce stable values of relative frequencies for each $\overline{\lambda}_j$ -- and each $a_i,\,b_k$ pair --, and all the steps of the derivation would almost ``literally''\footnote{Actual data is always subject to the loophole of finite statistics  \cite{pLar14}, so we can only expect experimental relative frequencies close to the ideal theoretical values. However, owing to the finiteness of the set, the classes $\overline{\lambda}_j$ would be ``exactly'' reproduced.} be realized with the use of classes $\overline{\lambda}_j\,,\,j\in\{1,\ldots,16\}$.
Thus, if the inequality is violated, one of the following hypothesis must be jettisoned
\begin{itemize}
\item LC; the RPCC fails and hidden variables as common causes do not exist.
\item MI; for some $(r,m)\neq(i,k)$, there exists $\overline{\lambda}_j$ such that $p(\overline{\lambda}_j|_{rm})\neq p(\overline{\lambda}_j|_{ik})$,\footnote{We are assuming that $p$ is appropriately defined to represent the distribution function of hidden variables' classes.} i.e., the relative frequency for some class of hidden variable depends on the settings which would block the transition from step (\ref{eq:s1}) to (\ref{eq:s2}) and the realization of (\ref{eq:cs}).
\end{itemize}
Then, either local causality or no conspiracy is violated.
\section{Joint Probabilities}\label{sec:JP}
Although this is somehow a different case and cannot be considered an orthodox position, it is worth mentioning in a Bell theorem's logical consistency discussion.
Most of these views seem to arise as a consequence of misinterpreting Fine's theorem.

Arthur Fine proved \cite{pFin82} that the BI holds if, and only if, a \emph{joint probability}(JP) $P(A_1,A_2,B_1,B_2)$ exists for the experiment's probabilities; this is commonly known as Fine's theorem.
He also proved that a LHV model is equivalent to the existence of a JP.
For our purposes, we call these results  Fine's theorem A and B, respectively.
{\small
\begin{eqnarray}
JP  &\leftrightarrow BI& \,(Fine's\,\, theorem\,A)\label{leq:FTA}\\
LHV &\leftrightarrow JP& \,(Fine's\,\, theorem\,B)\label{leq:FTB}\\
LHV &\rightarrow     BI& \,(Bell's\,\, theorem)\label{leq:BT}
\end{eqnarray}
}
The stance claiming that Fine has disproved Bell's theorem is unjustified since no rules of inference exist leading to any of the following implications
{\small
\begin{equation}\label{leq:fleq}
   \begin{array}{rrl}
       (\ref{leq:FTA}) &\rightarrow & \neg\,(LHV\rightarrow  BI)\\
       (\ref{leq:FTB}) &\rightarrow &\neg\,(LHV\rightarrow  BI)\\
       (\ref{leq:FTA})\wedge(\ref{leq:FTB}) &\rightarrow &\neg\,(LHV\rightarrow     BI)
   \end{array}
\end{equation}
}
Although, when put in a formal language, any implication in (\ref{leq:fleq}) is obviously false, the JP issue gave rise to some curious claims concerning the Bell theorem.

One of those asserts the BI is violated owed to the absence of a JP and that LHV plays no role in the argument.
Despite the previous statement being correct, it is also true that LHV implies the inequality and, as we have just proved in section \ref{sec:DERIV}, a JP plays absolutely no role in the argument.

An example may be useful to clarify previous the point.
By the same token, the expression $a(t)=a_0e^{-\alpha t}$ could not describe radioactive decay because it has been proven to describe the cooling of bodies; therefore, nuclear disintegration is not needed and plays no role in the argument.

Those kinds of arguments do not appear only in popular accounts and informal discussions but also in peer reviewed journals and books \cite{pGri20,pMuy86,bMuy02,pKhr19}.
The strategy consists of deriving the inequality from different hypotheses finding another context where a different interpretation applies, so nonlocality is irrelevant \cite{pSza94}.

A second slightly different claim asserts that a JP does not exist because it implies incompatible experiments; for instance, it is impossible to measure together $A_1$ and $A_2$, so a JP is meaningless and cannot exist, which would render LHV inconsistent according to (\ref{leq:FTB}) \cite{bMuy02,cKup12,pKup20}.

The fallacy, in this case, can be revealed with a counterexample. A JP could exist notwithstanding measurements incompatibility.
Suppose the hidden variable is obtained by throwing a dice at the source so that $\lambda\in\{1,2,3,4,5,6\}$, then the result is communicated through a classical channel to Alice and Bob.
In their respective laboratories, both experimenters toss a fair coin and evaluate the functions $A(a,\lambda)$ and $B(b,\lambda)$, where $a,b\in\{-1,+1\}$ with $H\equiv -1$ and $T\equiv +1$, given by
\begin{eqnarray}
A(a,\lambda) &=& a^\lambda\\
B(b,\lambda) &=& b^{\lambda+1}
\end{eqnarray}
For our purpose, we do not need perfect correlations; we only need a local hidden variable model.
The two possible values of $a$ and $b$ are clearly incompatible; a coin gives either $H$ or $T$ but not both.
As in the CHSH spin experiment where each party can measure only one direction but not both. We have
{\small
\begin{eqnarray}
\langle A_1B_1\rangle          &=&\sum_{i=1}^6 p(\lambda_i)1^\lambda 1^{\lambda+1}=\frac{1}{6}*6=1\\
\langle A_{-1}B_1\rangle       &=&\sum_{i=1}^6 p(\lambda_i)(-1)^\lambda 1^{\lambda+1}=\frac{1}{6}*0=0\\
\langle A_1B_{-1}\rangle       &=&\sum_{i=1}^6 p(\lambda_i)1^\lambda (-1)^{\lambda+1}=\frac{1}{6}*0=0\\
\langle A_{-1}B_{-1}\rangle    &=&\sum_{i=1}^6 p(\lambda_i)(-1)^\lambda (-1)^{\lambda+1}=\frac{1}{6}*(-6)=-1
\end{eqnarray}
}
Our model is local and, of course, satisfies the BI. For instance
{\small
\begin{equation}
-2\leq\langle A_1B_1\rangle - \langle A_{-1}B_1\rangle + \langle A_1B_{-1}\rangle + \langle A_{-1}B_{-1}\rangle=0\leq 2
\end{equation}
}
Hence, according to Fines's theorem A, a joint probability $P(A_1,A_{-1},B_1,B_{-1})$ does exist, although the experiments $A_{-1},A_{1}$ and $B_{-1},B_{1}$ are incompatible.

Thirdly, sometimes LHV is claimed to be too restrictive to describe the phenomenon because it only contemplates a single probability space, while it is well known in probability theory that some experiments require more than one probability space \cite{pNie11,pKhr09}.
According to this, the distribution function $p(\lambda)$ must depend on what they call ``the experimental context'', which amounts to including the setting parameters as explicit variables; $p_{ab}(\lambda)=p(\lambda|a,b)$.
This assertion is equivalent to no conspiracy violation, so physicists know about it, and they are well aware of the assumption.

Setting dependent distributions is also claimed to be justified by taking into account the measuring apparatuses' action. John Bell recognized this point and showed how to deal with it\cite{bBel71}. Th. M. Nieuwenhuizen and J. P. Lambare further discussed it \cite{pNie11,pLam17b}.
\section{Conclusions}
In section \ref{sec:ICDF}, we explained the irrelevance of the CFD-BI is twofold.
No meaningful comparison of the CFD-BI is possible either with the experimental results or with quantum mechanical predictions.
Furthermore, even not accepting CFD's inconsistent nature, the derivation of section \ref{sec:DERIV} clearly shows that CFD is utterly unnecessary, hence irrelevant concerning BI interpretation and consequences.
It is fair pointing out that H. P. Stapp's efforts to prove quantum mechanics nonlocality based on CR should not be confused with the application of CFD    to the BI. Stapp never claimed to have derived a falsifiable inequality.

A pragmatic scientist may adduce that, since CFD leads to the correct bound of the inequality, it is a valid and elegant principle that yields the BI as a trivial ``Fact''\cite{pGil14}.
However, this ``shut up and calculate'' dogmatic attitude is dangerous when it implies logical and methodological inconsistencies.
Such mistakes should not be dismissed as mere interpretational or philosophical problems.
We can only hope that medical and pharmaceutical researchers do not base their findings on CFD-like arguments, overlooking the rigid experimental and laboratory standards characterizing the scientific method.

We suspect that Bell's opinion of his inequality's counterfactual interpretation would have been the same he had of Von Neumann's 1932 impossibility proof:\emph{``When you translate them into terms of physical disposition they're nonsense. You may quote me on that. The proof of Von Neumann is not merely false but foolish''}\cite{mBel88}.

If the great Von Neumann committed such a silly mistake, we are all safe committing one.
However, that a whole community of scientists remained so uncritical is a worrisome sign of scientific conformism.
A few years later, Grete Germann pointed out Von Neumann's error \cite{pMer18}, but nobody listened.
More than thirty years later, after Bell's insistence \cite{pBel66}, physicists started to acknowledge the mistake.

After nearly fifty years, CFD's acceptance has not diminished.
It is directly used or assumed to underly the BI in articles published even by the most prestigious physical journals \cite{pGri20,pWol15,pWha20}.
Maybe it is time to pointing out the error explicitly and start making sense of the Bell inequality.
\section*{Acknowledgements}
The authors are grateful to Dr. Michael Hall for some enlightening discussions and useful bibliographic references.
\bibliography{zANLC}

\begin{thebibliography}{10}

\bibitem{pCHSH69}
J.F. Clauser, M.A. Horne, A.~Shimony, and R.A. Holt.
\newblock Proposed experiment to test local hidden-variables theories.
\newblock {\em Phys.Rev.Lett.}, 23:640--657, 1969.

\bibitem{wiki:cfd}
{Wikipedia contributors}.
\newblock Counterfactual definiteness --- {Wikipedia}{,} the free encyclopedia,
  2020.
\newblock [Online; accessed 3-January-2021].

\bibitem{pNor07}
T.~Norsen.
\newblock Against ``realism''.
\newblock {\em Found. Phys.}, 37:311--454, 2007.

\bibitem{pSta71}
H.~P. Stapp.
\newblock S-matrix interpretation of quantum theory.
\newblock {\em Phys. Rev. D}, 6 B:1303--1320, 1971.

\bibitem{pMau10}
Tim Maudlin.
\newblock What {B}ell proved: {A} reply to {B}laylock.
\newblock {\em American Journal of Physics}, 78:121, 2010.

\bibitem{pLau18a}
F.~Laudisa.
\newblock Counterfactual reasoning, realism and quantum mechanics: Much ado
  about nothing?
\newblock {\em Erkenn}, pages 1--16, 2018.

\bibitem{pSky82}
B.~Skyrms.
\newblock Counterfactual definiteness and local causation.
\newblock {\em Philosophy of Science}, 49 B:43--50, 1982.

\bibitem{pFor86}
M.~R. Forster.
\newblock Counterfactual reasoning in the {Bell-EPR} paradox.
\newblock {\em Philosophy of Science}, 53:133--144, 1986.

\bibitem{pDic93}
Michael Dickson.
\newblock Stapp's {T}heorem {W}ithout {C}ounterfactual {C}ommitments: {W}hy
  {I}t {F}ails {N}onetheless.
\newblock {\em Stud. Hist. Phil. Sci.}, 24:791--814, 1993.

\bibitem{pZuk14}
M.~Zukowski and C.~Brukner.
\newblock Quantum non-locality—it ain't necessarily so....
\newblock {\em Phys. A: Math. Theor.}, 47:424009, 2014.

\bibitem{pCza20}
Marek Czachor.
\newblock A {L}oophole of {A}ll ``{L}oophole-{F}ree'' {B}ell-{T}ype {T}heorems.
\newblock {\em Foundations of Science}, 25:971--985, 2020.

\bibitem{pLam20b}
J.~P. Lambare.
\newblock Comment on ``{A} {L}oophole of {A}ll ``{L}oophole-free''
  {B}ell-{T}ype {T}heorems''.
\newblock {\em Found Sci}, 2020.

\bibitem{pPer78}
A.~Peres.
\newblock Unperformed experiments have no results.
\newblock {\em American Journal of Physics}, 46:745--747, 1978.

\bibitem{pBla10}
G.~Blaylock.
\newblock The {EPR} paradox, {B}ell's inequality, and the question of locality.
\newblock {\em American Journal of Physics}, 78:111--120, 2010.

\bibitem{pGil14}
R.~D. Gill.
\newblock Statistics, causality and {B}ell's theorem.
\newblock {\em Statistical Science}, 29:512--528, 2014.

\bibitem{pBou17}
S.~Boughn.
\newblock Making sense of {B}ell's theorem and quantum nonlocality.
\newblock {\em Found. of Phys.}, 47:640--657, 2017.

\bibitem{pEPR35}
A.~Einstein, B.~Podolski, and Rosen N.
\newblock Can quantum-mechanical description of physical reality be considered
  complete?
\newblock {\em Phys.Rev.}, 47:777--780, 1935.

\bibitem{RPCC}
Christopher Hitchcock and Mikl\'{o}s R\'{e}dei.
\newblock {Reichenbach's Common Cause Principle}.
\newblock In Edward~N. Zalta, editor, {\em The {Stanford} Encyclopedia of
  Philosophy}. Metaphysics Research Lab, Stanford University, spring 2020
  edition, 2020.

\bibitem{pFra82}
B.~C. Van~Fraassen.
\newblock The {C}harybdis of realism: epistemological implications of {B}ell's
  inequality.
\newblock {\em Synthese}, 52:25--38, 1982.

\bibitem{pHal20}
Michael J.~W. Hall.
\newblock Does locality plus perfect correlation imply determinism?
\newblock {\em arXiv:2009.14223}, 2020.

\bibitem{bVai09}
Lev Vaidman.
\newblock {\em Counterfactuals in Quantum Mechanics}, pages 132--136.
\newblock Springer Berlin Heidelberg, Berlin, Heidelberg, 2009.

\bibitem{pHow85}
Don Howard.
\newblock Einstein on locality and separability.
\newblock {\em Studies in History and Philosophy of Science Part A},
  16(3):171--201, 1985.

\bibitem{pLam19}
J.~P. Lambare.
\newblock Bell inequalities, {C}ounterfactual {D}efiniteness and
  {F}alsifiability.
\newblock {\em arXiv:1911.00343 [quant-ph]}, 2019.

\bibitem{bBel71}
J.~S. Bell.
\newblock {\em Speakable and Unspeakable in Quantum Mechanics}, chapter
  Introduction to the hidden variable question, pages 36--37.
\newblock Cambridge University Press, Cambridge, 2004.

\bibitem{pGHZ90}
Daniel~M. Greenberger, Michael~A. Horne, Abner Shimony, and Anton Zeilinger.
\newblock Bell's theorem without inequalities.
\newblock {\em American Journal of Physics}, 58(12):1131--1143, 1990.

\bibitem{pHes04}
Karl Hess and Walter Philipp.
\newblock Breakdown of {B}ell's theorem for certain objective local parameter
  spaces.
\newblock {\em Proceedings of the National Academy of Sciences},
  101(7):1799--1805, 2004.

\bibitem{bAde01}
Guillaume Adenier.
\newblock {\em Foundations of Probability and Physics}, chapter Refutation of
  {B}ell's {T}heorem, pages 29--38.
\newblock World Scientific, 2001.

\bibitem{pJCh19}
J.~Christian.
\newblock Bell's {T}heorem versus {L}ocal {R}ealism in a {Q}uaternionic {M}odel
  of {P}hysical {S}pace.
\newblock {\em IEEE Access}, 7:133388--133409, 2019.

\bibitem{pJCh18}
J.~Christian.
\newblock Quantum correlations are weaved by the spinors of the {E}uclidean
  primitives.
\newblock {\em R. Soc open sci.}, 5:180526, 2018.

\bibitem{pDBa84a}
W.~{D}e Baere.
\newblock On the significance of {B}ell's inequality for hidden-variables
  theories.
\newblock {\em Lettere Al Nuovo Cimento}, 39(11):234--238, 1984.

\bibitem{pEbe77}
P.~Eberhard.
\newblock Bell's {T}heorem without {H}idden {V}ariables.
\newblock {\em Nuov. Cim.}, 38 B:75--79, 1977.

\bibitem{pNie09}
Th.~M. Nieuwenhuizen.
\newblock Where {B}ell {W}ent {W}rong.
\newblock {\em AIP Conference Proceedings}, 1101(1):127--133, 2009.

\bibitem{pLar14}
Jan-{\AA}ke Larsson.
\newblock Loopholes in {B}ell inequality tests of local realism.
\newblock {\em Journal of Physics A: Mathematical and Theoretical},
  47(42):424003, oct 2014.

\bibitem{pLam17a}
J.~P. Lambare.
\newblock On the {CHSH} form of {B}ell's inequalities.
\newblock {\em Found. Phys.}, 47:321--326, 2017.

\bibitem{pWig70}
Eugene~P. Wigner.
\newblock On {H}idden {V}ariables and {Q}uantum {M}echanical {P}robabilities.
\newblock {\em American Journal of Physics}, 38:1005, 1970.

\bibitem{pFin82}
A.~Fine.
\newblock Hidden variables, joint probability, and the {B}ell inequalities.
\newblock {\em Phys. Rev. Lett.}, 48:291--295, 1982.

\bibitem{pGri20}
Robert Griffiths.
\newblock Nonlocality claims are inconsistent with {Hi}lbert-space quantum
  mechanics.
\newblock {\em Physical Review A}, 101:022117, 2020.

\bibitem{pMuy86}
W.~M.~De Muynck.
\newblock The {B}ell {I}nequalities and their {I}rrelevance to the {P}roblem of
  {L}ocality in {Q}uantum {M}echanics.
\newblock {\em PHYSICS LETTERS}, 114A(2):65--67, 1986.

\bibitem{bMuy02}
W.~M. Muynck.
\newblock {\em Foundations of Quantum Mechanics, and Empiricist Approach},
  chapter The {B}ell inequality in quantum mechanics, pages 471--478.
\newblock Kluwer Academic Publishers, 2002.

\bibitem{pKhr19}
Andrei Khrennikov.
\newblock Get {R}id of {N}onlocality from {Q}uantum {P}hysics.
\newblock {\em Entropy}, 21(8):806, 2019.

\bibitem{pSza94}
L\'{a}szl\'{o}~E. Szab\'{o}.
\newblock On the real meaning of {B}ell's theorem.
\newblock {\em International Journal of Theoretical Physics}, 33:191--197,
  1994.

\bibitem{cKup12}
M.~Kupczynski.
\newblock Entanglement and quantum nonlocality demystified.
\newblock {\em AIP Conf. Proc.}, (1508):253, 2012.

\bibitem{pKup20}
M.~Kupczynski.
\newblock Is the {M}oon {T}here {I}f {N}obody {L}ooks: {B}ell {I}nequalities
  and {P}hysical {R}eality.
\newblock {\em Frontiers in Physics}, 8:273, 2020.

\bibitem{pNie11}
Th.~M. Nieuwenhuizen.
\newblock Is the contextuality loophole fatal for the derivation of {B}ell
  inequalities?
\newblock {\em Found. Phys.}, 41:580--591, 2011.

\bibitem{pKhr09}
Andrei Khrennikov.
\newblock Nonlocality as well as rejection of realism are only sufficient (but
  non-necessary!) conditions for violation of {B}ell's inequality.
\newblock {\em Information Scienes}, 179(5):492--504, 2009.

\bibitem{pLam17b}
J.~P. Lambare.
\newblock On {N}ieuwenhuizen's treatment of contextuality in {B}ell's theorem.
\newblock {\em Found. Phys.}, 47:1591--1596, 2017.

\bibitem{mBel88}
J.~S. Bell.
\newblock Interview {J}ohn {B}ell.
\newblock {\em Omni}, 10(8):84--92, May 1988.

\bibitem{pMer18}
David~N. Mermin and R.~Schack.
\newblock Homer {N}odded: {V}on {N}eumann's {S}urprising {O}versight.
\newblock {\em Found Phys}, 48:1007--1020, May 2018.

\bibitem{pBel66}
J.~S. Bell.
\newblock On the problem of hidden variables in quantum mechanics.
\newblock {\em Reviews of Modern Physics}, 38:447--452, 1966.

\bibitem{pWol15}
S.~Wolf.
\newblock Nonlocality without counterfactual reasoning.
\newblock {\em Phys. Rev. A}, 92:052102, 2015.

\bibitem{pWha20}
K.~B. Wharton and N.~Argaman.
\newblock Colloquium: Bell's theorem and locally mediated reformulations of
  quantum mechanics.
\newblock {\em Rev. Mod. Phys.}, 92:021002, May 2020.

\end{thebibliography}
\end{document}